\documentclass[10pt,journal]{IEEEtran}

\usepackage{setspace}
\usepackage{graphicx}
\usepackage{amstext}
\usepackage{algorithm}
\usepackage{algorithmic}
\usepackage{mathrsfs}
\usepackage{amssymb}
\usepackage{amsmath}
\usepackage{bm}
\allowdisplaybreaks[4]
\usepackage{epstopdf}
\usepackage{multicol}
\usepackage{stfloats}
\usepackage{url}
\usepackage{color}
\usepackage{enumerate}
\usepackage{amsmath}
\usepackage{breqn}
\usepackage{bbm}
\usepackage{cite}
\usepackage{subfig}
\usepackage{caption}
\usepackage{enumitem}
\usepackage{mathrsfs}
\usepackage{multirow}
\allowdisplaybreaks[4]

\begin{document}

	\title{Enhancing Communication Efficiency of Semantic Transmission via Joint Processing Technique}

	\author{Xumin Pu, \emph{Member, IEEE}, Tiantian Lei, Wanli Wen, \emph{Member, IEEE},  and Qianbin Chen, \emph{Senior Member, IEEE}
		
		\thanks{This work was supported in part by the National Natural Science Foundation of China under Grants 62201101 and 61701062, in part by the Science and Technology Research Program of Chongqing Municipal Education Commission (KJQN202100649), in part by the Project funded by China Postdoctoral Science Foundation under Grant 2022M720020, in part by the Natural Science Foundation of Chongqing, China  under Grant cstc2021jcyj-msxmX0458, in part by the Chongqing Technology Innovation and Application Development Special Key Project under Grant  CSTB2022TIAD-KPX0059, and in part by the open research fund of National Mobile Communications Research Laboratory, Southeast University under Grant 2022D06.  \emph{(Corresponding author: Wanli Wen.)}}
		\thanks{X. Pu, T. Lei, and Q. Chen are with the School of Communications and Information Engineering, Chongqing University of Posts and Telecommunications, Chongqing 400065, China (e-mail: puxm@cqupt.edu.cn, S210101066@stu.cqupt.edu.cn, and chenqb@cqupt.edu.cn).}
		\thanks{W. Wen is with the School of Microelectronics and Communication Engineering, Chongqing University, China and also with the National Mobile Communications Research Laboratory, Southeast University, Nanjing, China (wanli\_wen@cqu.edu.cn).}
		
	}	
	


	\maketitle
	
	\begin{abstract}
		This work presents a novel semantic transmission framework in wireless networks, leveraging the joint processing technique.  Our framework enables multiple cooperating base stations to efficiently transmit semantic information to multiple users simultaneously. To enhance the semantic communication efficiency of the transmission framework, we formulate an optimization problem with the objective of maximizing the semantic spectral efficiency of the framework and propose a low-complexity dynamic semantic mapping and resource allocation algorithm. This algorithm, based on deep reinforcement learning and alternative optimization, achieves near-optimal performance while reducing computational complexity. Simulation results validate the effectiveness of the proposed algorithm, bridging the research gap and facilitating the practical implementation of semantic communication systems.
	\end{abstract}


	\IEEEpeerreviewmaketitle
	\vspace{-0.5cm}
	\section{Introduction}
	Amid the proliferation of wireless applications and the exponential growth of data traffic, conventional communication paradigms that transmit multimedia content (e.g., text, images, and videos) in wireless networks encounter a formidable challenge due to the scarcity of available communication resources. This situation necessitates a paradigm shift from conventional to semantic communications. Semantic communications demonstrate substantial potential in mitigating communication resource shortages by prioritizing the transmission of the intended meaning of multimedia content, garnering extensive attention from industry and academia \cite{9475174}. Nonetheless, transmitting the semantic information of a significant volume of multimedia content can pose challenges to semantic communication efficiency, such as transmission rate or spectral efficiency \cite{2022arXiv220101389Q}. Consequently, how to enhance semantic communication efficiency has been regarded as a significant obstacle in the practical implementation of semantic communication systems.
	
	Recently, a significant focus has been on developing efficient semantic communication systems, as highlighted in several notable references \cite{9953095, 9797984, 9763856, 9747455, 10122232, 10000686, 9472999}. Specifically, for text transmission, the authors in \cite{9953095} and \cite{9797984} proposed two different resource allocation algorithms to enhance the semantic transmission rate. In a similar vein, \cite{9763856} introduced a joint resource allocation and semantic mapping algorithm to improve semantic spectral efficiency. Note that these solutions only rely on traditional numerical optimization methods, which may face challenges when applied to dynamic and time-varying wireless channels. To address these challenges, \cite{9747455} proposed the utilization of deep reinforcement learning (DRL) to maximize the overall semantic similarity of recovered texts by users and developed a DRL-based resource allocation algorithm to achieve this objective. For image transmission, \cite{10122232} devised a DRL-based joint semantic compression ratio and resource allocation algorithm to optimize semantic efficiency. Furthermore, \cite{10000686} examined a DRL-based approach for joint semantic information allocation and resource allocation, with the goal of minimizing semantic transmission latency. In the context of video transmission, \cite{9472999} proposed a DRL-based semantic information allocation algorithm that specifically targets the minimization of semantic rate distortion in video frames.  
	
	In the realm of conventional communications, the joint processing (JP) technique has been effectively utilized to enhance communication efficiency \cite{5706317}. By enabling coordinated cooperation among multiple base stations (BSs) to serve users, the JP technique has shown significant improvements in communication efficiency for conventional communication systems. Building upon these advancements, it becomes intriguing to investigate whether similar benefits can be extended to semantic transmission in wireless networks. Surprisingly, this aspect has yet to be explored in the existing literature in the field of semantic communications. Therefore, the potential benefits of JP in enhancing communication efficiency for semantic transmissions remain unknown. 
	
	In this paper, we would like to bridge the aforementioned research gap by establishing a novel semantic transmission framework in wireless networks, utilizing the JP technique. Our contributions can be summarized as follows:
	\begin{itemize}
		\item We propose a novel semantic transmission framework that leverages the JP technique in a wireless network. This framework enables multiple cooperating BSs to efficiently transmit semantic information to multiple users simultaneously.
		\item We formulate an optimization problem to maximize the semantic spectral efficiency (SSE) for all users. This problem is inherently challenging due to its non-convex and mixed-integer nature, which makes it generally classified as NP-hard.
		\item We develop a low-complexity dynamic semantic mapping and resource allocation algorithm based on DRL and the alternative optimization method. This algorithm effectively optimizes the SSE while mitigating computational complexity, enabling practical implementation in real-world scenarios.
	\end{itemize}
	\begin{figure}[H]
		\begin{center}		
			\includegraphics[width=0.47\textwidth]{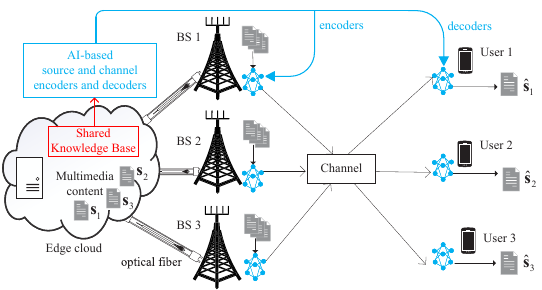}\\
		\end{center}
		\caption{System model with $K=N=3$. }
		\label{figsystemmodel}
		\vspace{-3.5mm}
	\end{figure}
	\section{System Model}
	As illustrated in Fig.~\ref{figsystemmodel}, we consider a semantic transmission framework in a wireless network,  which consists of one edge cloud, $N$ BSs represented by $\mathcal{N} \triangleq \{1,2,\cdots, N\}$, and $K$ users represented by $\mathcal{K} \triangleq \{1,2,\cdots, K\}$. The edge cloud stores a shared knowledge base and $K$ multimedia files, such as text, images, audio, and video, where the $k$-th file, denoted by $\mathbf{s}_k$, represents the desired content for user $k \in \mathcal{K}$. Based on the shared knowledge base, similar to \cite{9398576}, we consider training an AI-based joint source and channel codec in the edge cloud.  Once trained, the codec will be deployed to the BSs and users, respectively. This deployment can enable the BSs to encode and the users to  decode the semantic information of the multimedia files. When user $k$ requests $\mathbf{s}_k$ from the edge cloud, all BSs will retrieve file $k$ via backhaul, encode the semantic information of $\mathbf{s}_k$ using the source and channel encoders, and simultaneously transmit the information to user $k$ over the wireless channel using the JP technique. After receiving the semantic information from the channel, user $k$ decodes the information via the integrated channel and source decoders and obtains its content, denoted by $\hat{\mathbf{s}}_k$.  We examine a discrete-time system where time is segmented into $T$ slots, denoted as a set $\mathcal{T} \triangleq \{1,2,\cdots, T\}$. It is assumed that each slot has an equal length, and within a single slot, the channel state remains static, varying between slots. In the following, we focus on the system operation in the $t$-th slot.
	
	The semantic transmission framework under consideration is assumed to be text-based, i.e., the multimedia files stored in the edge cloud are text files. However, it is crucial to emphasize that the analytical and optimization framework presented in this paper can be applied to semantic communication systems supporting the transmission of other types of multimedia. Let us consider that the file $\mathbf{s}_k$ consists an average of $L_k$ words. In slot $t$, after passing the AI-based source and channel encoders, $\mathbf{s}_k$ is mapped to a semantic symbol vector $\mathbf{x}^{t}_k$ of length $\tau^{t}_k L_k $. Here, $\mathbf{x}^{t}_k$ depicts the semantic information of $\mathbf{s}_k$ and $\tau^{t}_k $ denotes the average number of semantic symbols used for each word in $\mathbf{s}_k$. We stipulate that $\tau^{t}_k $ has an upper bound, denoted by $\Gamma$, i.e.,  $\tau^{t}_k$ must satisfy the following constraint
	{\setlength\abovedisplayskip{0pt}
		\setlength\belowdisplayskip{4pt}
		\begin{align}\label{eqtau}
		\tau^{t}_{k} \in\{1,2,\cdots, \Gamma\} , \quad  k \in {\mathcal{K}}.
		\end{align}}Define ${\boldsymbol\tau^{t}}\triangleq(\tau^{t}_{k})_{k \in {\mathcal{K}}}$ as the semantic mapping design for all users in $\mathcal{K}$ in the $t$-th slot.

	
	We consider that each BS is equipped with $M$ transmit antennas, while each user has a single receive antenna. In the $t$-th slot, let $\mathbf{w}^{t}_{n,k} \in \mathbb{C}^{M\times 1}$ represent the transmit beamformer for user $k$ on the $n$-th BS with $\mathbb{C}$ standing for the complex number. Denote $\mathbf{w}^{t}_k \triangleq (\mathbf{w}^{t}_{n,k})_{n\in\mathcal{N}}$ as the beamforming design of user $k$ and $\mathbf{w}^{t}\triangleq (\mathbf{w}^{t}_k)_{k\in\mathcal{K}}$  the beamforming design of all users in $\mathcal{K}$. To accommodate the power resource constraint at BS $n$, we impose the following restriction on vector $\mathbf{w}^{t}$
	{\setlength\abovedisplayskip{1pt}
		\setlength\belowdisplayskip{4pt}
		\begin{align}\label{eqwnk}
		\sum_{k \in {\mathcal{K}}}  (\mathbf{w}^{t}_{n,k})^{\rm H} \mathbf{w}^{t}_{n,k} \leq P_n, \quad  n \in {\mathcal{N}},
		\end{align}}where $P_n$ denotes the maximum transmit power of BS $n$ and $(\cdot)^{\rm H}$ stands for the conjugate transpose operator. Consider that all BSs in set $\mathcal{N}$ employ the JP technique to cooperatively transmit semantic information to all users in set $\mathcal{K}$ in each time slot, utilizing the same bandwidth.
	Let $\mathbf{h}^{t} \triangleq (\mathbf{h}^{t}_k)_{k\in\mathcal{K}}$ denote the channel state of all users in $\mathcal{K}$ and $\mathbf{h}^{t}_k \triangleq (\mathbf{h}^{t}_{n,k})_{n\in\mathcal{N}}$ the channel state of user $k$, where $\mathbf{h}^{t}_{n,k} \in \mathbb{C}^{M \times 1}$ represents the channel coefficient from BS $n$ to user $k$. 
	Then, we can calculate the received signal-to-interference-plus-noise-ratio (SINR) of the $k$-th user in the $t$-th slot as
	{\setlength\abovedisplayskip{2pt}
		\setlength\belowdisplayskip{4pt}
		\begin{align}\label{eqpower}
		\gamma_{k} (\mathbf{w}^{t}_k; \mathbf{h}^{t}_k)=\frac{\left| \sum_{n \in {\mathcal{N}}} (\mathbf{h}^{t}_{n,k})^{\rm H} \mathbf{w}^{t}_{n,k}\right|^2 }{\sum_{i \in \mathcal{K} \backslash \{k\} } \left|\sum_{n \in \mathcal{N}}(\mathbf{h}^{t}_{n,k})^{\rm H} \mathbf{w}^{t}_{n,i}\right|^2 +\sigma^2}.
		\end{align}}Here, $\sigma^2$ denotes the noise power. To ensure a successful information decoding, we have the following constraint
	{\setlength\abovedisplayskip{2pt}
		\setlength\belowdisplayskip{4pt}
		\begin{align}\label{eqSINR}
		\gamma_k(\mathbf{w}^{t}_k; \mathbf{h}^{t}_k) \geq \gamma_{\text {th }} ,  \quad  k \in {\mathcal{K}},
		\end{align}}where $\gamma_{\text {th }}$ is a SINR threshold.

	
	We evaluate the performance of the semantic transmission framework using the SSE metric proposed in \cite{9763856}. The SSE metric measures the rate at which semantic information can be transmitted effectively within a unit of bandwidth, expressed in semantic units per second per hertz (suts/s/Hz). This metric provides an analytical characterization of the communication efficiency in text-based semantic communication systems. Specifically, the SSE of user $k$ in the $t$-th slot can be defined~as
	{\setlength\abovedisplayskip{0pt}
		\setlength\belowdisplayskip{4pt}	
		\begin{align}\label{eqRk}
		R^{t}_k = \frac{ I_k}{\tau^{t}_k L_k} \xi^{t}_k.
		\end{align}}
	Here, $I_{k}$ is the average amount of semantic information of ${\textbf s}_{k}$, and $\xi^{t}_k\in [0,1]$ represents the semantic similarity between $\mathbf{s}_k$ and $\hat{\mathbf{s}}^{t}_k$ in the
	$t$-th slot. Assuming that the AI-based source and channel encoders/decoders are implemented using the Transformer model, we can calculate $\xi^{t}_k$ as
	{\setlength\abovedisplayskip{3pt}
		\setlength\belowdisplayskip{4pt}
		\begin{align}\label{eqksikdef}
		\xi^{t}_k\triangleq\frac{{\rm{B}}({\mathbf s}_{k}) {\rm{B}}(\hat{\mathbf s}^{t}_k)^{\rm{T}}}{\|{\rm{B}}({\mathbf s}_{k})\|\|{\rm{B}}(\hat{\mathbf s}^{t}_k)\|}.
		\end{align}}
	Here, $\rm{B}(\cdot)$ represents the sentence-bidirectional encoder representations from the Transformer model \cite{2022arXiv220101389Q1}. It is worth noting that semantic similarity indirectly reflects communication reliability. A high semantic similarity between $\mathbf{s}_k$ and $\hat{\mathbf{s}}^{t}_k$ indicates reliable communication, whereas a low similarity suggests potential unreliability in the communication process. To ensure communication reliability, $\xi^{t}_k$ should satisfy the following constraint
	{\setlength\abovedisplayskip{1pt}
		\setlength\belowdisplayskip{4pt}
		\begin{align}\label{eqksik}
		\xi^{t}_k \geq \xi_{\text {th }} ,  \quad  k \in {\mathcal{K}}.
		\end{align}}
	Here, $\xi_{\text {th }}$ denotes a semantic similarity threshold.	
	
	\section{Problem Formulation and Transformation}
	
	Based on (\ref{eqRk}), we can calculate the total SSE of all users in $\mathcal{K}$ in the
	$t$-th slot as $R^{t} \triangleq \sum_{k\in\mathcal{K}} R^{t}_k$. We would like to maximize $R^{t}$ under the constraints in (\ref{eqtau}), (\ref{eqwnk}), (\ref{eqSINR}), and (\ref{eqksik}). Specifically, we  can formulate the following problem. 
	
	\textit{Problem 1 (SSE Maximization):}
	{\setlength\abovedisplayskip{2pt}
		\setlength\belowdisplayskip{4pt}
		\begin{equation*}\label{probSSEMax}
		{\setlength\abovedisplayskip{3pt}
			\setlength\belowdisplayskip{3pt}
			\begin{aligned}
			\max _{\boldsymbol{\tau}^{t},\mathbf{w}^{t}} & \quad  R^{t}\nonumber\\
			\mathrm{s.t.}\;
			&\mbox{(\ref{eqtau}), (\ref{eqwnk}), (\ref{eqSINR}), (\ref{eqksik}).}
			\end{aligned}}
		\end{equation*}}
	Unfortunately, solving Problem 1 directly is unattainable, as the semantic similarity $\xi^{t}_k$ involved in the total SSE $R^{t}$ lacks an explicit form.
	To tackle this issue, inspired by \cite{9953095}, we employ the generalized logistic function to approximate $\xi^{t}_k$ as $\Tilde{\xi}_k(\tau^{t}_k,\mathbf{w}^{t}_k; \mathbf{h}^{t}_k)$, i.e.,
	{\setlength\abovedisplayskip{2pt}
		\setlength\belowdisplayskip{2pt}
		\begin{align}\label{eqapproxik}
		\xi^{t}_k \approx  \Tilde{\xi}_k(\tau^{t}_k,\mathbf{w}^{t}_k; \mathbf{h}^{t}_k) \triangleq a_{\tau^{t}_{k}}+\frac{b_{\tau^{t}_{k}}}{1\!+\!e^{-c_{\tau^{t}_{k}}  (\gamma_{k}(\mathbf{w}^{t}_k; \mathbf{h}^{t}_k)+d_{\tau^{t}_{k}}\!)}}. 
		\end{align}}Here, $a_{\tau^{t}_{k}}$, $b_{\tau^{t}_{k}}$, $c_{\tau^{t}_{k}}$, and $d_{\tau^{t}_{k}}$ are parameters that can be determined using the criterion of nonlinear least squares \cite{9953095}.
	Based on this, we can approximate the total SSE $R^{t}$ as follows 
	{\setlength\abovedisplayskip{0pt}
		\setlength\belowdisplayskip{4pt}
		\begin{align}\label{eqR}
		R^{t} \approx \Tilde{R}(\boldsymbol{\tau}^{t}, \mathbf{w}^{t}; \mathbf{h}^{t}) \triangleq \sum_{k\in\mathcal{K}} \Tilde{R}_k(\tau^{t}_k, \mathbf{w}^{t}_k; \mathbf{h}^{t}_k).
		\end{align}}
	Here, $\Tilde{R}_k(\tau^{t}_k, \mathbf{w}^{t}_k; \mathbf{h}^{t}_k)$ is an approximation of $R^{t}_k$ in (\ref{eqRk}), which can be calculated by replacing $\xi^{t}_k$ in (\ref{eqksikdef}) with $\Tilde{\xi}_k(\tau^{t}_k,\mathbf{w}^{t}_k; \mathbf{h}^{t}_k)$ in (\ref{eqapproxik}),~i.e.,
	{\setlength\abovedisplayskip{2pt}
		\setlength\belowdisplayskip{4pt}
		\begin{align}\label{eqRk0}
		\Tilde{R}_k(\tau^{t}_k, \mathbf{w}^{t}_k; \mathbf{h}^{t}_k) \triangleq \frac{ I_k}{\tau^{t}_k L_k}  \Tilde{\xi}_k(\tau^{t}_k,\mathbf{w}^{t}_k; \mathbf{h}^{t}_k).
		\end{align}}Based on (\ref{eqapproxik}), (\ref{eqR}), and (\ref{eqRk0}), we can approximately transform Problem 1 into the following problem.
	
	\textit{Problem 2 (Approximation Problem of Problem 1):}
	{\setlength\abovedisplayskip{3pt}
		\setlength\belowdisplayskip{3pt}
		\begin{equation*}\label{probAppx}{\setlength\abovedisplayskip{3pt}
			\setlength\belowdisplayskip{3pt}
			\begin{aligned}
			\max _{\boldsymbol{\tau}^{t},\mathbf{w}^{t}} & \quad   \Tilde{R}(\boldsymbol{\tau}^{t}, \mathbf{w}^{t}; \mathbf{h}^{t}) \nonumber\\
			\mathrm{s.t.}\;
			&\mbox{(1), (2), (4),}\nonumber 
			\end{aligned}}
		\end{equation*}}
	{\setlength\abovedisplayskip{-3pt}
		\setlength\belowdisplayskip{-1pt}
		\begin{align}\label{eqRk0}
		\qquad \qquad \qquad \qquad	\quad  \Tilde{\xi}_k(\tau^{t}_k,\mathbf{w}^{t}_k; \mathbf{h}^{t}_k) \geq \xi_{\text {th }} ,  \quad  k \in {\mathcal{K}}
		\end{align}} 
	
	Compared to Problem 1, Problem 2 explicitly presents the objective function and all constraints. Therefore, our focus shifts to solving Problem 2 rather than Problem 1.  Problem 2 is a  mixed integer nonlinear programming problem.  The main challenge in solving it lies in the presence of the integer variable $\boldsymbol{\tau}^{t}$.   Conventional optimization algorithms require iterative adjustments of the integer variable to find an optimal solution, which has exponential time complexity in the worst case and is impractical for dynamic optimization in the presence of rapidly fading channels. To address this issue,  we propose a DRL-based dynamic semantic mapping and resource allocation (DSMRA) algorithm in the following section. The DSMRA algorithm is capable of achieving polynomial time complexity in solving Problem 2.
	\vspace{-0.4cm}
	\section{Problem Solution}
	
\begin{figure}
		\begin{center}		\includegraphics[width=0.4\textwidth]{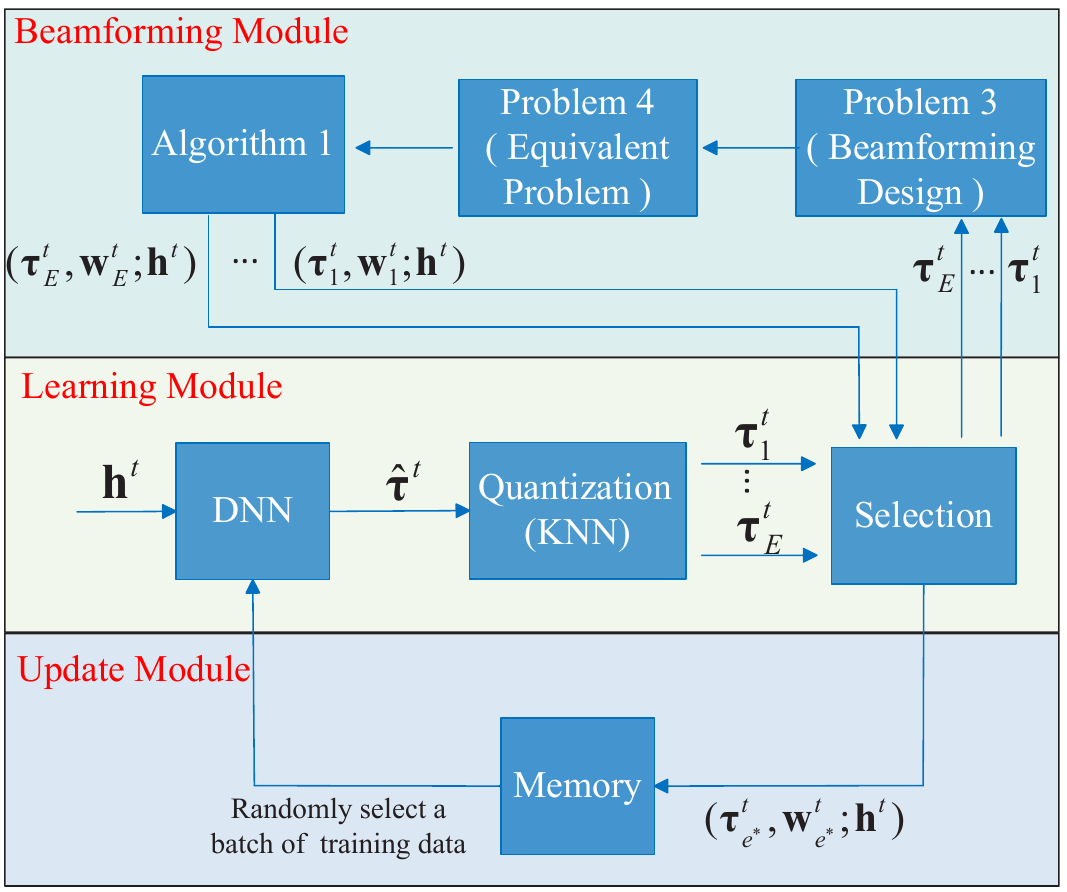}\\
		\end{center}
		\caption{The schematics of the proposed DSMRA algorithm. }
		\label{figDSMRA}
		\vspace{-6mm}
	\end{figure}

	The schematic of our proposed DSMRA algorithm is presented in Fig.~\ref{figDSMRA}, which consists of three main modules: the learning module, the beamforming module, and the update module. The learning module leverages a deep neural network (DNN) to swiftly generate a promising mapping action for Problem 2 as soon as the channel realization is available at the beginning of slot $t$. The beamforming module is responsible for generating the corresponding beamforming policy, which determines the beamforming vectors for each user based on the mapping action obtained from the learning module. The update module is designed to enhance the learning capability of the learning module by periodically refreshing the training data, which ensures that the DNN remains up to date and can adapt to changing channel conditions, improving the overall performance of the algorithm. In the following sections, we will provide a detailed explanation of each module. 
	
	\vspace{-0.4cm}
	\subsection{Learning Module} 
	
	This module comprises three sub-modules: the DNN sub-module, the quantization sub-module, and the selection sub-module. In slot $t$, the DNN sub-module takes the channel state $\mathbf{h}^{t}$ as input and produces a continuous mapping action $\hat{\boldsymbol{\tau}}^t$. Mathematically, this process can be represented as $\hat{\boldsymbol{\tau}}^t = f_{\theta^t}(\mathbf{h}^t)$, where $f_{\theta^t}(\cdot)$ denotes the DNN with parameters $\theta^t$, and $\hat{\boldsymbol{\tau}}^t \triangleq (\hat{\tau}^{t}_k)_{k\in\mathcal{K}}$ with $\hat{\tau}^{t}_k \in [1, \Gamma]$. The quantization sub-module quantizes $\hat{\boldsymbol{\tau}}^t$ into $E$ possible mapping designs using the K-Nearest Neighbors (KNN) method, where $E$ can be any integer within $ [1,\Gamma^K]$. This process can be expressed as $\{\boldsymbol{\tau}^{t}_e\}_{e\in\mathcal{E}} = Q_E(\hat{\boldsymbol{\tau}}^t)$, where $Q_E(\cdot)$ denotes the KNN operation, $\mathcal{E} \triangleq \{1,2,\cdots, E\}$, and $\boldsymbol{\tau}^{t}_e \triangleq (\tau_{e,k}^t)_{k\in\mathcal{K}}$ with $\tau_{e,k}^{t} \in \{1,2,\cdots, \Gamma\}$, for all $e\in\mathcal{E}$. The selection sub-module takes $\boldsymbol{\tau}^{t}_e$ to the beamforming module (see Section IV-B) to determine the corresponding beamforming design $\mathbf{w}^t_e$ for all $e\in\mathcal{E}$ and selects an optimal mapping and beamforming pair, denoted by $(\boldsymbol{\tau}^{t}_{e^*}, \mathbf{w}^t_{e^*}; \mathbf{h}^t)$, according to $e^* \triangleq \arg\max_{e \in \mathcal{E}} \Tilde{R}(\boldsymbol{\tau}^{t}_e, \mathbf{w}^{t}_e; \mathbf{h}^{t})$. Note that, $(\boldsymbol{\tau}^{t}_{e^*}, \mathbf{w}^t_{e^*}; \mathbf{h}^t)$ will be stored by the update module for refreshing the training data (see Section IV-C). 
	\vspace{-0.3cm}
	\subsection{Beamforming Module}

	Mathematically, this module is to solve Problem 2 when $\boldsymbol{\tau}^t$ is determined.  Specifically, given $\boldsymbol{\tau}^t=\boldsymbol{\tau}^t_e$, Problem 2 reduces to the following beamforming design problem.
	
	\textit{Problem 3 (Beamforming Design):}
	{\setlength\abovedisplayskip{2pt}
		\setlength\belowdisplayskip{4pt}
		\begin{equation*}\label{probRA}
		\begin{aligned}
		\max_{\mathbf{w}^t} & \quad  \Tilde{R}(\boldsymbol{\tau}^t, \mathbf{w}^t; \mathbf{h}^t)  \nonumber\\
		\mathrm{s.t.}\;
		&\mbox{(2), (4), (11).}
		\end{aligned}
		\end{equation*}}
	\begin{algorithm}[htb] 
		\caption{Algorithm for Solving Problem 4 with Given Mapping Design}
		\begin{algorithmic}[1]
			\STATE {Set $j :=  0$ and $\boldsymbol{\eta}^t_{0} := \gamma_{\rm th}\mathbf{I}$ for all $k\in\mathcal{K}$ with $\mathbf{I}$ denoting a vector with all components being $1$.}
			\STATE \textbf{repeat}
			\STATE {\quad  Obtain $\mathbf{w}^t_{j+1}$ by solving Problem 4 with given $\boldsymbol{\eta}^t_{j}$ using an interior-point method.}
			\STATE {\quad  Check the constraints (14). If they are not satisfied, stop the algorithm and set $\mathbf{w}^t_e := \mathbf{w}^t_{j+1}$, $\Tilde{R}(\boldsymbol{\tau}^t_{e}, \mathbf{w}^t_{e};  \mathbf{h}^t) = 0$}. Otherwise, obtain $\boldsymbol{\eta}^t_{j+1}$ by solving Problem 4 with given $\mathbf{w}^t_{j+1}$ using an interior-point method.
			\STATE { \quad Update $j := j+1$.}
			\STATE \textbf{until} {stopping criterion is met.}
			\STATE {Set $\mathbf{w}^t_e := \mathbf{w}^t_{j+1}$ and obtain $\Tilde{R}(\boldsymbol{\tau}^t_{e}, \mathbf{w}^t_{e};  \mathbf{h}^t)$}.
		\end{algorithmic}\label{algforprobeqRA}
		\vspace{-1mm}
	\end{algorithm}
	Problem 3 is a non-convex continuous optimization problem. It is worth noting that for some values of $\boldsymbol{\tau}^t_e$, the problem may become infeasible. In such cases, we set the objective function value to zero. To solve Problem 3, we first introduce a new optimization variable, $\eta^t_{k}$, and additional constraint, i.e., 
	{\setlength\abovedisplayskip{1.5pt}
		\setlength\belowdisplayskip{4pt}	
		\begin{align}\label{eqetak}
		\gamma_k(\mathbf{w}^t_k; \mathbf{h}^t_k) \geq  \eta^t_{k} ,  \quad  k \in {\mathcal{K}}.
		\end{align}}
	Then, we equivalently convert the constraints in (\ref{eqSINR}) and (11)~to
	{\setlength\abovedisplayskip{-2pt}
		\setlength\belowdisplayskip{1pt}
		\begin{align}
		&\eta^t_k \geq \gamma_{\text {th }} ,  \quad  k \in {\mathcal{K}}, \label{eqetakth} \\
		&a_{\tau^t_{k}}+\frac{b_{\tau^t_{k}}}{1+e^{-c_{\tau^t_{k}}  (\eta^t_{k}+d_{\tau^t_{k}})}}\geq \xi_{\text {th }},  \quad  k \in {\mathcal{K}}, \label{eqxik0th}
		\end{align}}and the objective function $\Tilde{R}(\boldsymbol{\tau}^t, \mathbf{w}^t; \mathbf{h}^t)$ to
	{\setlength\abovedisplayskip{0pt}
		\setlength\belowdisplayskip{2pt}
		\begin{align*}
		\Tilde{R}^{'}(\boldsymbol{\tau}^t, \mathbf{w}^t, \boldsymbol{\eta}^t; \mathbf{h}^t) = \sum_{k \in {\mathcal{K}}} \frac{I_{k}}{\tau^t_{k} L_{k}} \left(a_{\tau^t_{k}}\!+\!\frac{b_{\tau^t_{k}}}{1\!+\!e^{-c_{\tau^t_{k}}  \!(\eta^t_{k}+d_{\tau^t_{k}}\!)}}\right),
		\end{align*}}where ${\boldsymbol\eta}^t\triangleq(\eta^t_{k})_{k \in {\mathcal{K}}}$. As a result, Problem 3 is equivalently transformed into the following 
	problem.
	
	\textit{Problem 4 (Equivalent Problem of Problem 3):}
	{\setlength\abovedisplayskip{1pt}
		\setlength\belowdisplayskip{2pt}
		\begin{equation*}\label{probeqRA}
		\begin{aligned}
		\max_{\mathbf{w}^t, \boldsymbol{\eta}^t} \quad & \Tilde{R}^{'}(\boldsymbol{\tau}^t, \mathbf{w}^t, \boldsymbol{\eta}^t; \mathbf{h}^t)  \nonumber\\
		\mathrm{s.t.}\;
		&\mbox{(2), (12), (13), (14).}
		\end{aligned}
		\end{equation*}}
	Problem 4 is a bi-convex optimization problem, which means that it is convex with respect to one variable while the other variable is fixed. Specifically, Problem 4 is convex w.r.t. $\mathbf{w}^t$ if $\boldsymbol{\eta}^t$ is fixed, and it is convex w.r.t. $\boldsymbol{\eta}^t$ if $\mathbf{w}^t$ is fixed. This property can be helpful in solving the optimization problem using iterative methods that alternate between optimizing each variable while keeping the other fixed. The details for solving Problem 4 is summarized in Algorithm~\ref{algforprobeqRA}. 
	\renewcommand{\algorithmicrequire}{\textbf{Input:}} 
	\renewcommand{\algorithmicensure}{\textbf{Output:}}
	\begin{algorithm}[htb]
		\caption{The DSMRA Algorithm}
		\label{alg:Framwork}
		\textbf{Input:} The channel state $\mathbf{h}^t$. \\
		\textbf{Output:} Optimal mapping design ${\boldsymbol{\tau}^{t}_{e^*}}$ and beamforming design $\mathbf{w}^t_{e^*}$ for each slot $t$.
		\begin{algorithmic}[1]
			\STATE Initialize the DNN with random parameters $\theta^1$ and
			empty memory.
			\label{code:fram:extract}			,
			\label{code:fram:trainbase}
			\FOR{$t=1,2,\dots,T$}
			\STATE Generate a continuous mapping action $\hat{\boldsymbol{\tau}}^t$.
			\STATE Quantize $\hat{\boldsymbol{\tau}}^t$ into $E$ possible mapping designs $\{\!\boldsymbol{\tau}^{t}_e\!\}_{e\in\mathcal{E}}$.
			\FOR{$e=1,2,\dots,E$}
			\STATE \quad Obtain $ \Tilde{R}(\boldsymbol{\tau}^{t}_e, \mathbf{w}^{t}_e; \mathbf{h}^{t})$ by using Algorithm~\ref{algforprobeqRA}.
			\ENDFOR
			\STATE Select an optimal mapping and beamforming pair  
			\quad $(\boldsymbol{\tau}^{t}_{e^*}, \mathbf{w}^t_{e^*}; \mathbf{h}^t)$ and add it to the memory.
			\STATE \textbf{if} {$t$ mod $\Delta$ = 0}
			\STATE \quad Randomly select a batch of training data from \\ \quad memory to enhance the learning module.
			\STATE \textbf{end if}
			\ENDFOR
		\end{algorithmic}
	\end{algorithm}

	We now analyze the convergence and computational complexity of Algorithm~\ref{algforprobeqRA}. Since both subproblems in Problem 4 are convex, the interior-point method is employed to find their optimal solutions. Thus, Algorithm~\ref{algforprobeqRA} will converge to a stationary point of Problem 4. The complexity of Algorithm~\ref{algforprobeqRA} depends on solving these two subproblems. If the interior-point method is utilized, the complexity of Algorithm~\ref{algforprobeqRA} is $O\left(J((K N)^{3.5} M^3+3K^{2}) \log \epsilon^{-1}\right)$ \cite{9234527}, where $\epsilon$ is the required accuracy of the duality gap termination, and $J$ represents the number of iterations required for the algorithm to converge. 
	\begin{figure}\vspace{-3mm}
		\begin{center}		\includegraphics[width=0.35\textwidth]{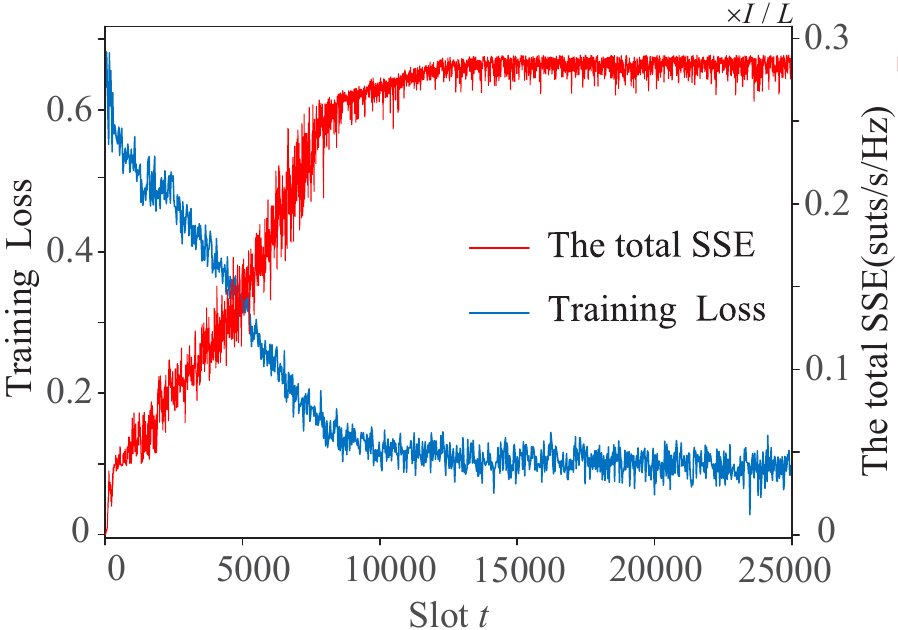}\\
		\end{center}
		\caption{The convergence for the DSMRA algorithm. }
		\label{figloss}
		\vspace{-6mm}
	\end{figure}\vspace{-0.5cm}

	\begin{figure*}[h]	
	\begin{minipage}{0.32\linewidth}
		\vspace{3pt}
		\centerline{\includegraphics[width=\textwidth]{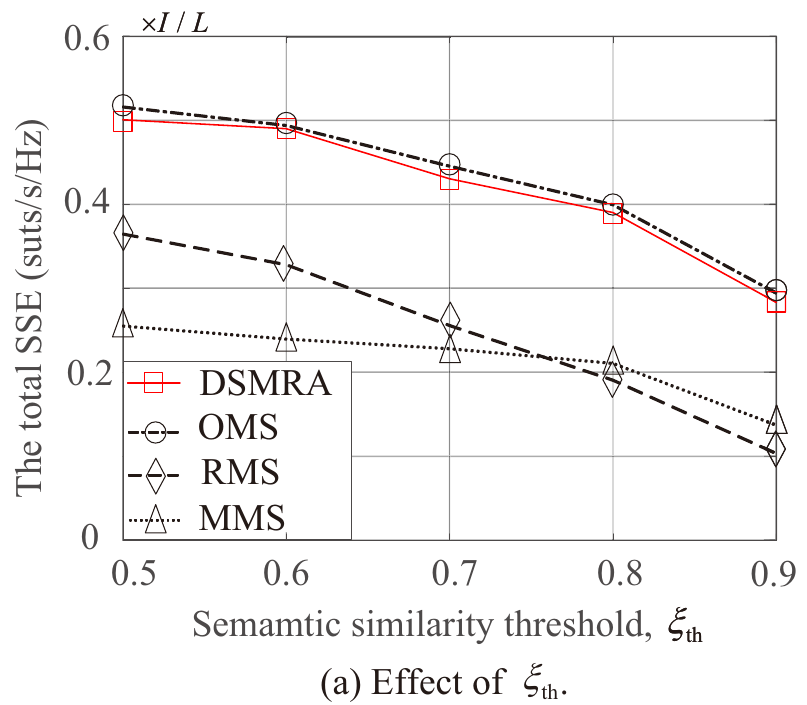}}
	\end{minipage}
	\begin{minipage}{0.32\linewidth}
		\vspace{3pt}
		\centerline{\includegraphics[width=\textwidth]{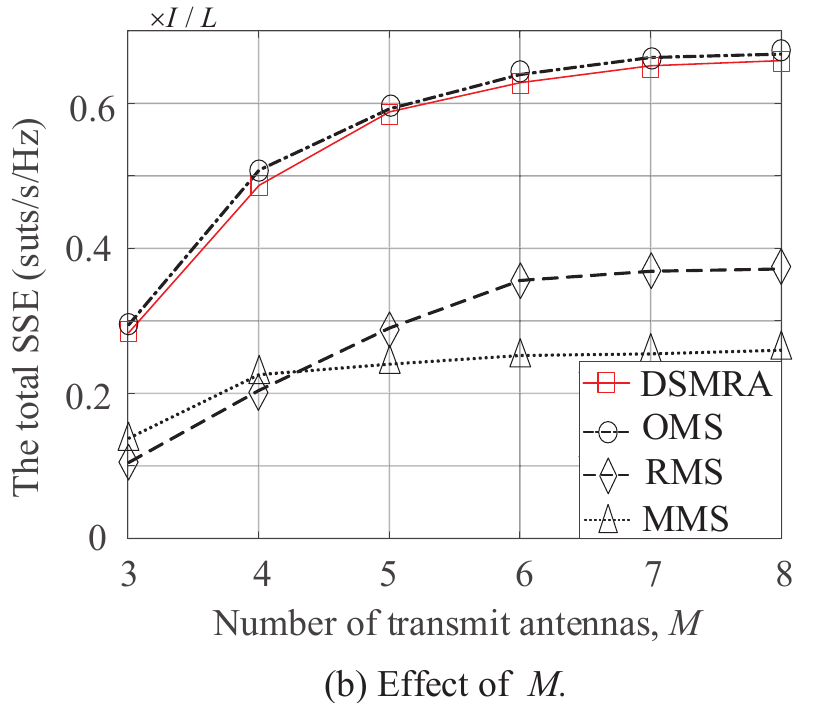}}
	\end{minipage}
	\begin{minipage}{0.32\linewidth}
		\vspace{3pt}
		\centerline{\includegraphics[width=\textwidth]{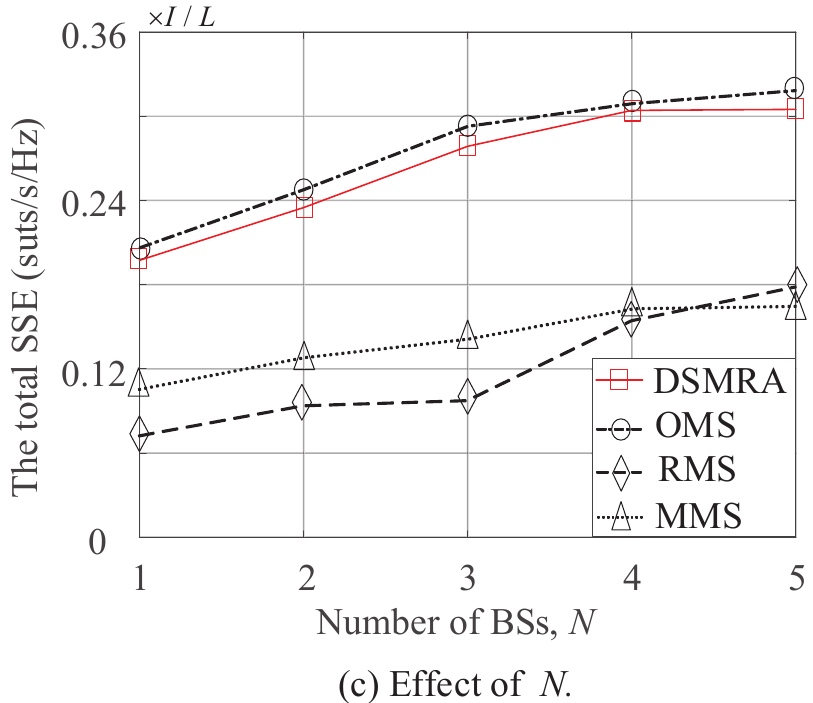}}
	\end{minipage}	
	\caption{Performance evaluation of the proposed algorithm and baselines.}
	\label{figperformence}
	\vspace{-6mm}
\end{figure*}

	\subsection{Update Module}  
	This module utilizes the experience replay technique to construct a periodically refreshed training dataset for the DNN. Initially, a limited-capacity memory is established, which is empty. Each new training data is added to the memory, replacing the oldest data if the memory is full. Then, every $\Delta$ slots, a batch of training data is randomly selected from the memory to train the DNN. The cross-entropy is used as the training loss, and the Adam algorithm is applied to minimize it. This iterative process gradually enhances the learning capability of the module. Since the limited-capacity memory is utilized, the complexity of batch updates is reduced and the correlation between training samples is lowered, thereby accelerating the convergence speed. Once the training loss converges, the module is considered successfully trained and capable of generating promising mapping actions for Problem 2 when the channel realization is available. The overall DSMRA algorithm is summarized in Algorithm~\ref{alg:Framwork}.
		
	\vspace{-0.4cm}
	\section{Simulation Results}
	

	
	Except when noted otherwise, the simulation parameters are specified as $M=3$, $P_n=30$ dBm, for all $n\in\mathcal{N}$, $\sigma^2=-83.98$ dBm, $\xi_{\text {th }}=0.9$, $\gamma_{\text {th}}=3.7$ dBm, $\Gamma=11$ (symbols/word), and $I_{k}=I$, $L_{k}=L$, for all $k \in {\mathcal{K}}$. In addition, we consider a setup with $N=3$ BSs and $K=3$ users randomly distributed within a 500 meters radius disk. The channel states of all users are chosen according to the distance-dependent path-loss model, defined as $\mathrm{PL}(d) = 128.1 + 37.6\lg(d)$ dB, where $d$ is the distance measured in kilometers, incorporating $8$ dB log-normal shadowing and a Rayleigh component. Our DSMRA algorithm is based on a four-layer neural network, with layers comprising $K$, 120, 80, and $K$ neurons, in that order. The training rate is set at 0.001, memory size at 1024, $\Delta=10$, and $E=K$. It is compared to three baseline schemes to evaluate the effectiveness of the proposed DSMRA algorithm. 
	\begin{itemize}
		\item  \textbf{Optimal Mapping Scheme (OMS)}: This scheme exhaustively searches among $\Gamma^K$ possible mapping designs to find an optimal one under the constraints in (\ref{eqwnk}), (\ref{eqSINR}), and (11). Due to the vast search space in (\ref{eqtau}), finding the optimal solution could be time-consuming.
		
		\item \textbf{Random Mapping  Scheme (RMS)}: This scheme randomly chooses a mapping design from (\ref{eqtau}). The beamforming design is then deduced via Algorithm~\ref{algforprobeqRA}.
		
		\item \textbf{Maximum Mapping Scheme (MMS)}: This scheme consider $\tau_k = \Gamma$, for all $ k\in {\mathcal{K}}$. The beamforming design is then deduced via Algorithm~\ref{algforprobeqRA}.
	\end{itemize}


	Fig.~\ref{figloss} shows the convergence for our DSMRA algorithm.  Initially, the algorithm exhibits high loss and low total SSE due to an unenhanced learning module that yields random mapping designs. However, through the reinforcement learning process, the learning module uses past training data to continually improve itself. As the advent of one slot after another, the loss gradually plateaus and stabilizes around~0.1,  while the total SSE stabilizes around~0.28. The minor fluctuations are largely due to the randomness inherent in the training data, indicating successful training of the learning module.
	
	Fig.~\ref{figperformence} shows the performance comparison of the proposed DSMRA algorithm with other baseline schemes. Several observations are: First, as the semantic similarity threshold rises, the total SSE of all schemes decreases. This is due to the increased utilization of semantic symbols per word. Second, with an increase of either $M$ or $N$, each scheme's total SSE rises. This is a consequence of having more antennas which provide additional spatial freedom, thus reducing user interference with appropriate beamformer selection and consequently decreasing the average number of semantic symbols per word. Third, our proposed DSMRA algorithm, leveraging reinforcement learning for optimal mapping strategy, performs comparably to the OMS scheme and significantly surpasses the other two. This superior performance results from DSMRA's learning capability, which the RMS and MMS schemes lack, to map accurately based on the problem's characteristics.

	\section{Conclusion}
	
	In this work, we introduced a novel semantic transmission framework in wireless networks, leveraging the JP technique. This framework enables multiple cooperating BSs to simultaneously transmit semantic information to multiple users. We formulated an optimization problem to maximize the SSE metric and proposed a low-complexity algorithm based on DRL and alternative optimization methods. The algorithm achieves near-optimal performance while mitigating computational complexity, demonstrating its practical viability. Future work may involve further refinement of the algorithm and investigating its performance in diverse network scenarios.


\bibliographystyle{IEEEtran}

\end{document}